# Digital laser frequency control and phase stabilization loops for a high precision space-borne metrology system

G. Hechenblaikner, V. Wand, M. Kersten,
K. Danzmann, A. Garcia, G. Heinzel, M. Nofrarias, and F. Steier

**Abstract—We report on the design, implementation and characterization of fully digital control loops for laser frequency stabilization, differential phase-locking and performance optimization of the optical metrology system on-board the LISA Pathfinder space mission. The optical metrology system consists of a laser with modulator, four Mach-Zehnder interferometers, a phase-meter and a digital processing unit for data analysis. The digital loop design has the advantage of easy and flexible controller implementation and loop calibration, automated and flexible locking and resetting, and improved performance over analogue circuitry. Using the practical ability of our system to modulate the laser frequency allows us to accurately determine the open loop transfer function and other system properties. Various noise sources and their impact on system performance are investigated in detail.**

## I. INTRODUCTION

LASER stabilization has been of crucial importance to meet the requirements for timing and frequency standards of many modern applications ranging from ultra-precise optical clocks [1,2] to accurate long-distance measurements for a detailed mapping of the earth gravitational field [3,4] or, as in this paper, a prototype of a space-borne gravitational wave detector [5,6]. The lasers used in such precision measurements can be made to run stably at a certain frequency by locking them to a reference which may either be an atomic transition in a frequency-modulation spectroscopy configuration [7,8] or a stable cavity [9,10] in a Pound-Drever-Hall configuration [11,12]. Both methodologies are in fact quite similar: whereas in the first case frequency modulation (FM) spectroscopy is used to probe a narrow atomic resonance, in the second case the resonance of a cavity mode is probed and the laser is locked to the cavity using a Pound-Drever Hall lock. These methods do not allow to easily frequency modulate the laser as can be done in the scheme described in this paper. Here, in some aspects similar to what was demonstrated in [13], we use an unbalanced (unequal path-length) Mach Zehnder interferometer to frequency-lock the laser to a fringe appearing in the interference pattern. Whereas this method does not give the ultra-narrow linewidths obtained from the cavity-locking schemes (e.g. see [9,10]), it still allows us to obtain linear spectral densities of $\sim 1\,kHz/\sqrt{Hz}$ at $1\,mHz$ and additionally offers the advantage of frequency-tunability as it does not require to lock at specific, discrete transmission peaks. Instead, the laser frequency can be changed over many GHz and the laser locked to any desired frequency. In our scheme we obtain optimal frequency stability from using an ultra-stable optical bench and a highly accurate phase-meter. The optical bench is insensitive to temperature fluctuations and path length variations through careful design and choice of build-materials, which are limiting factors for its relative frequency stability. The bench design also allows in a simple and straight-forward extension to add additional interferometers with balanced path-lengths but otherwise similar to the one used for frequency stabilization.

However, the actual purpose of the metrology is to measure position and attitude of two free-floating test-masses over a large dynamic range with picometer and nanorad accuracy, respectively. Optimal stabilization of laser frequency and optical path-length control constitute the basis for achieving this goal.

This paper is structured as follows:

In **section II** we give a brief overview over the LISA Pathfinder mission.

In **section III** we discuss the central elements of the Optical Metrology System (OMS). In particular the optical bench, phase-meter and algorithmic processing are described which are central to understanding the loop functionality and performance.

In **section IV** we discuss the individual loops in detail and

Manuscript received October 2010. This work was supported in part by the DLR (Deutsches Zentrum für Luft- und Raumfahrt).

G. Hechenblaikner, M. Kersten, and V. Wand are with EADS Astrium (e-mail: Gerald.Hechenblaikner[at]astrium.eads.net, Vinzenz.Wand[at]astrium.eads.net, Michael.Kersten[at]astrium.eads.net, ).

K. Danzmann, A. Garcia, G. Heinzel, M. Nofrarias, and F. Steier are with the Albert Einstein Institut, Hannover, Germany (email:Karsten.Danzmann[at]aei.mpg.de, Antonio.Garcia[at]aei.mpg.de, Gerhard.Heinzel[at]aei.mpg.de, Miquel.Nofrarias[at]aei.mpg.de , Frank.Steier[at]aei.mpg.de).

present a model of the loop architecture. Theoretical noise models are established for the laser- and optical path-length stabilization loops which are related to the actually measured noise spectra in the next section.

In **section V** we describe the measurements of the open loop transfer function. We show how accurate values for actuator gains and loop delay are determined from fits of the transfer function to the model of section 4 and investigate the limiting noise sources for the control loops. Finally we analyze the impact of various noise sources on the overall accuracy of the position and attitude measurements.

## II. THE LISA PATHFINDER MISSION

LISA Pathfinder [5,6] is the technological precursor to the actual LISA (Laser Interferometer Space Antenna) mission to detect gravitational waves through interferometric measurements of space-time distortions over large distances in space [5]. LISA consists of an equilateral triangular constellation of three space-craft with 5 million kilometres side-length and two floating test-masses at each vertex point that act as end-mirrors of interferometers sensitive to position fluctuations. Some of the LISA technology cannot be tested on ground. Therefore the Pathfinder mission was designed to test hardware with a design suitable for LISA and to verify important measurement principles.

To facilitate costs and testing, the Pathfinder mission employs only a single space-craft containing an interferometer with two floating test-masses whose separation is measured through an interferometer. In Pathfinder the residual acceleration of the test-masses is required to be measured with a sensitivity greater than $3 \times 10^{-14} m\, s^{-2} / \sqrt{Hz}$ [6], which implies a resolution of the test-mass position better than $6.4 \times 10^{-12} m / \sqrt{Hz}$ in the measurement band between 3 mHz and 30 mHz. The exact expression for the required linear spectral density (LSD) of the position measurement $x$, an indication of the measurement accuracy, is given by Equation 1:

$$LSD(x) \leq 6.4 \times 10^{-12} \sqrt{1 + \left(\frac{f}{3mHz}\right)^{-4}} \; m / \sqrt{Hz}, \qquad (1).$$

$1mHz \leq f \leq 30\,mHz$

In this paper we describe the measures taken to achieve this sensitivity through careful design of laser control and optical path-length stabilization loops

The Optical Metrology System (OMS) acts as an inertial sensor by measuring the position deviation (and, through calculation, the acceleration) of free test-masses. It is one of the central subsystems of the Pathfinder payload and essential in achieving the required mission performance. In science mode its sensor output together with the output from capacitive sensors serves as error-signal to the Drag Free Attitude and Control System (DFACS) [14] which controls the space-craft using micro-Newton thrusters as actuators. While the motion of the two test-masses and their attitude is constrained in some directions, the remaining six degrees of freedom (in a very simplistic model) allow a nearly free evolution of the test-mass motion. DFACS has to steer the space-craft in such a way that the test-masses remain well-centred in the confining electrode housings.

## III. THE OPTICAL METROLOGY SYSTEM

General information on the OMS is given in [15] and a detailed discussion on calibration and commissioning is found in [16]. Note that the measurement results of [16] are important in the present discussion of this paper and shall be referred to on several occasions. Here we restrict ourselves to a brief overview over the system functionality and only provide those details which are necessary for a proper understanding of the laser loop functionality.

The Optical Metrology System consists of

    A. Laser Assembly
    B. Optical bench interferometer
    C. Phase-meter
    D. Data Management Unit

Figure 1 depicts a schematic of laser assembly and optical bench with photo-diodes. The phase-meter and DMU, which process the photo-diode signals, are not shown.

### A. The Laser Assembly

The Laser assembly consists of a laser unit and two modulators. At the core of the laser unit is a Nd:YAG NPRO (non planar ring oscillator) emitting about 40mW linear polarized light at 1064 nm with tuning inputs for laser frequency and power control. At its output the beam is split into "reference" and "measurement" beams which pass through the optical modulators shifting their relative frequencies by the heterodyne frequency $f_{het} = 1 kHz$. These accousto-optic modulators are driven by high-powered radio-frequency (RF) signals which are synchronized to a common clock shared with the phase-meter and data-management unit. Through variation of the RF-amplitudes the modulators also serve as actuators for the fast power stabilization control loops which derive their error signal from the single element photo-diodes Pow1 and Pow2 (see Figure 1), respectively. Additionally, the laser modulator contains two piezo-actuated retro-reflectors which stabilize the relative optical path-length difference between the two beams by locking the phase $\varphi_R$ of the reference interferometer (see next section).

### B. The optical bench

Measurement and reference beam are transmitted to the optical bench through optical fibres. The optical bench consists of four non polarizing Mach-Zehnder interferometers. Two of them are the "science" interferometers, named "x1" and "x12". The x1 interferometer measures the absolute position and attitude of test-mass 1 with respect to the bench on photo-diode PD1. The x12 interferometer measures the relative position and attitude between the two test-masses on photo-diode PD12. To this end the measurement beam is reflected from test-mass 1 in the former and from both test-masses in the latter interferometer. For the on-ground testing described in this article the floating test-masses are replaced by dummy mirrors located at the nominal test-mass position outside the bench.

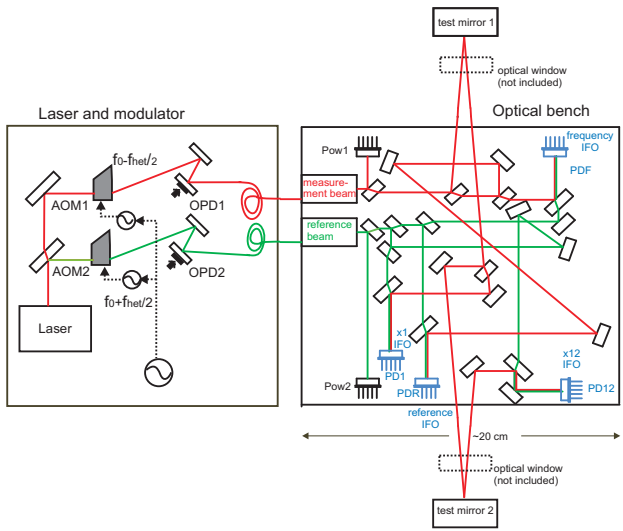

Figure 1: Schematic of the laser modulator, and the optical bench (drawn to scale). The latter is comprised of four independent heterodyne Mach-Zehnder interferometers, referred to as "x1", "x12", "frequency" and "reference" interferometer with the suffix 1, 12, F, and R, respectively. Optical windows of the inertial sensor housing (not included) are presented by the dotted rectangles. The beam paths of the x1 and the x12 interferometers are marked by the solid lines. The beam paths of the reference and frequency interferometers are denoted by the dotted lines.

The laser modulator contains two accousto-optic modulators, AOM1 and AOM2, which shift the frequencies of the beams relative to another by $f_{het}$. The optical path-lengths of measurement and reference beam are stabilized by the actuators OPD1 and OPD2.

The reference interferometer "R" provides a reference phase $\varphi_R$ that is subtracted from the phase-measurements of the other three interferometers ("x1","x12","F") so that optical path length variations occurring in the common optical path of all four interferometers cancel. The most significant contributions to path-length variations are attributed to optical fibres, fibre couplers and laser modulators, all of which are shared by the four interferometers and therefore cancel when subtracting the reference phase. Owing to the quasi-monolithic properties of the Zerodur® optical bench and its small coefficient of thermal expansion the optical path-length variations occurring on the optical bench are negligible. Therefore, subtracting the reference phase $\varphi_R$ from the detected phases $\varphi_i$ of the other three interferometers should ideally result in very stable relative output phases $\psi_i = \varphi_i - \varphi_R$, largely unaffected by path-length variations. However, cross-couplings of RF signals going to the accousto-optic modulators result in small optical sidebands, which in turn lead to significant optical path-length noise unless the path-length differences are stabilized (for a detailed account of this effect see [17]). To this end an optical path length difference stabilization loop is operated with the reference phase $\varphi_R$ serving as error signal.

The frequency interferometer "F" has deliberately mismatched optical path-lengths (including the optical fibre path), with one optical path-length exceeding the other by $L \sim 38\,cm$, so that laser frequency fluctuations $\delta\nu_{las}$ are converted into detectable phase fluctuations $\delta\varphi$ given by:

$$\delta\varphi = 2\pi \cdot \delta\nu_{las} \cdot L / c \qquad (2).$$

The frequency interferometer output serves as error signal to the frequency stabilization loops which are described in more detail further down. The other three interferometer have well-balanced optical path-lengths (to within 1 cm) so that laser frequency noise is suppressed and –for perfectly balanced optical paths- would disappear altogether.

### C. Phase-meter and photo-diodes

The interference signals are detected by quadrant photo-diodes, and each interferometer is equipped with a nominal and a redundant diode. Whereas the nominal diodes are clearly labelled, the redundant diodes are not further discussed in this paper and remain unmarked in Figure 1. The phase-meter samples the signals at the sampling frequency $f_{samp} = 50\,kHz$ and then performs a discrete Fourier transform at a repetition rate of $f_{FFT} = 100\,Hz$. In addition to the DC-component it only retains a single bin of the discrete frequency grid which corresponds to the beat frequency of the heterodyne signal $f_{het} = 1\,kHz$. The bin contains a complex amplitude vector with information on the relative phase between the two interfering beams and on the amplitude of the beat signal from which the contrast can be inferred. We found in previous investigations that the phase-meter is capable of measuring the longitudinal phase with an accuracy better than $10\,\mu rad / \sqrt{Hz}$ in the measurement band [16].

### D. Data Management unit and Digital Signal Processing

The data are transmitted at a rate of $100\,Hz$ from the phase-meter to the data-management unit (DMU). Due to constraints in the communication bandwidth, the DMU down-samples the data from $100\,Hz$ to $10\,Hz$ by application of a moving average filter. This only applies to "science data" relating to distance measurements whereas the DMU controlled digital loops still operate at the full bandwidth of $100\,Hz$. The down-sampled data are then transferred to the experimental control computer where they are stored and archived. Alternatively, without prior down-sampling, short periods (~16 seconds) of $100\,Hz$ real-time data can be written into the DMU memory and later retrieved for analysis.

The application software running on the DMU processes the incoming data, retrieves the amplitude and phase information, applies a phase-tracking algorithm, and calculates the attitude and position of the test-masses. It also calculates the interference contrasts and monitors the data quality of each channel.

## IV. STABILIZATION AND CENCELLATION LOOPS

### A. General information on the loops

In this section we briefly discuss the various types of control loops which are essential in achieving the measurement accuracy required in Equation 1. Note that the requirement for the accuracy of a position measurement relates to an equivalent requirement for the accuracy of a phase measurement, as phase $\psi$ relates to position $x$ through

$$x = \frac{\lambda}{4\pi \cos \delta} \psi ,\qquad (3)$$

where $\delta = 4.5\,\deg$ is the acute incidence angle of the measurement beam on the test-mass. Therefore laser frequency fluctuations couple into the position measurement as described by Equation 2 and must be reduced by a frequency control loop.

Similarly, optical path-length variations, although they should cancel by interferometer design, couple into the position measurement through higher order effects related to unwanted optical sidebands induced by the optical modulators [17]. To this end the Optical Path-Length Difference (OPD) control loop is designed to suppress OPD noise by stabilizing the relative optical path-lengths between measurement and reference beam. This corresponds essentially to a phase-lock-loop of the reference phase $\varphi_R$, the difference phase between the two beams measured in the reference interferometer. Note that although control of only one actuator in either measurement or reference beam path would be sufficient, the OPD loop controls actuators OPD1 and OPD2 located in both beams in a pull-push mode. This provides some degree of redundancy should one actuator fail.

The power stabilization loop is designed to counteract power fluctuations in the two beams. As a consequence of the phase-meter processing, the power fluctuations couple into the phase measurements of all interferometers as an additional contribution to the phase noise. As power fluctuations are common mode to all interferometers and their respective interference signals, one might tend to assume that they cancel when the reference phase $\varphi_R$ is subtracted from the phase $\varphi_i$ of another interferometer to yield the relative phase $\psi_i$ which is then used in the further processing. However, as we will see further down, depending on the value of the relative phase $\psi_i$, common mode noise generally does not completely cancel in the subtraction. The fast power control loop is designed to suppress the power fluctuations. It operates at ~ 50 kHz bandwidth, is internally controlled by the laser assembly and reduces laser power fluctuations at the heterodyne frequency from $\sim 1 \times 10^{-5} / \sqrt{Hz}$ to $\sim 1 \times 10^{-6} / \sqrt{Hz}$ in the measurement band ($1\,mHz - 30\,mHz$). An additional slow power loop, acting on the laser diode current, is designed to offload the actuators of the fast power loop, avoid saturation of the latter, and to define a basic setpoint for the laser output power.

The digital control laws for the various loops were all chosen based on defined accuracy requirements as well as open loop noise measurements.

### B. The loop controller

The loop controllers, one for each loop, are implemented digitally in the application software running on the DMU. Their implementation allows us to choose an operating point comprised of a static setpoint $\Delta x$ and a dynamic modulation $\delta x(t_i)$, as described in Figure 2. In its general form the controller is represented by a fifth order infinite-impulse response (IIR) filter, but for reasons of simplicity we use controllers up to third order only. The digital representation of the control law is given by $H_{discrete} = \sum B_i z^{-i} / \sum A_i z^{-i}$, where $A_i, B_i$ are the filter coefficients stored in the software and $z = e^{j\omega}$. For reasons of practicality and ease of modelling, we shall usually refer to its analogue equivalent which is obtained from the digital controller through a Tustin transformation.

The open loop gain is defined as the ratio between controller input $x$ and control error $e$ : $x/e = H \cdot G$. The closed loop gain is defined as the ratio between controller modulation $\delta x$ and controller error $e$ : $e/\delta x = 1/(1 + HG)$.

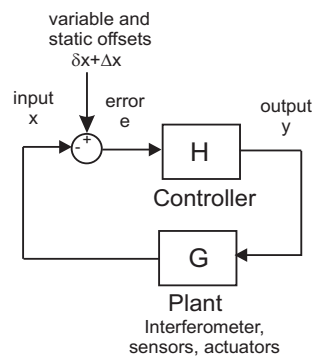

Figure 2: Schematics of the loop controller. Static $\Delta x$ and dynamic $\delta x(t_i)$ setpoints are added to the controller input $x$ to obtain the error $e = \Delta x + \delta x - x$. The controller acts on $e$ and produces the output $y$.

### C. The laser frequency control loop

A schematic of the frequency control loop is given in Figure 3a. In our discussion of the various loop components we shall start with the laser output of frequency $\nu_{las} = c/\lambda$. The laser frequency and its noise $n_{las}$ are converted into a fluctuating phase $\varphi_F$ through the action of the frequency interferometer, which –according to Equation 2- applies a gain factor of $G_1 = 2\pi L/c$.

More noise $n_{lfo}$ is added through noise picked up in the optical fibres, mechanical micro-vibrations and temperature fluctuations before the signal is detected by the photo-diodes and routed to the phase-meter. When the analog-to-digital converter (ADC) samples the signal it adds quantization noise, transition noise and distortion to the signal which we shall collectively refer to as "ADC noise". The previously analogue interferometer phases $\varphi_F, \varphi_R$ are thereby converted with unity gain ($G_2 = 1$) into digital phases. In this process noise sources acting on the signal amplitude, that is ADC-noise $n_{ADC}$ and laser power fluctuations $n_{pow}$, are converted from amplitude noise to phase-noise. The ADC noise contribution has been measured systematically before [15] and found to be $n_{ADC} \sim 10\,\mu rad/\sqrt{Hz}$ .

An approximate analytic derivation yields that the linear spectral density of the laser power noise ($\delta P/P$) transfers proportionally into an equivalent phase-noise through the action of the phase-meter: $n_{pow} = (\delta P/P)\,rad/\sqrt{Hz}$ .For our laser system the power fluctuations were measured to be $n_{pow} \sim 1\,\mu rad/\sqrt{Hz}$ when the power stabilization loop is locked.

Finally, the reference phase is subtracted from the phase of the frequency interferometer to obtain the relative phase $\psi_F = \varphi_F - \varphi_R$ , thereby removing (or at least strongly suppressing) all common mode noise between interferometers.

At this point the total loop delay $\tau_{freq}$ is symbolically represented which includes delays incurred due to finite sampling rate (at $100\,Hz$), data- and command-communication, and an extra delay incurred by a serial to MIL-STD-1553 interface converter between phase-meter and data-management unit.

In the next step static and dynamic offsets, $\Delta x$ and $\delta x(t_i)$, respectively, can optionally be added to the inverted signal to yield the error $e$ that serves as input to the fast frequency controller with a transfer function $G_{3f} = (0.79s - 158)/s$ . The controller output goes to the fast actuator, a piezoelectric transducer (PZT) in the laser head, with a gain of $G_{4f} \sim 2\,MHz/V$ as well as to the slow frequency controller with a transfer function $G_{3s} = (2.5 \cdot 10^{-5}s - 5 \cdot 10^{-3})/s$ . The output of the slow controller is routed to the slow actuator (temperature controller in laser head) with a gain of $G_{4s} \sim 500\,MHz/V$ and a response time of $T \sim 0.5\,s$ . Both, the fast and the slow actuator, add some noise to their output, which is estimated to be smaller than $1\,\mu rad/\sqrt{Hz}$ and $100\,\mu rad/\sqrt{Hz}$ , respectively. Considering that this noise is suppressed by the total loop gain, the actuator noise is sufficiently small to be neglected.

Taking into account all gain stages and transfer functions we obtain the following model for the open loop transfer function

$$G_{tot} = \frac{0.79s - 158}{s}G_1\left[G_{4f} + G_{4s}\frac{2.5\cdot10^{-5}s - 5\cdot10^{-3}}{s}\frac{1}{0.5s+1}\right]e^{s\tau}, \qquad (4)$$

where we left the actuator gains $G_{4f}, G_{4s}$ and the total loop delay $\tau_{freq}$ as free parameters.

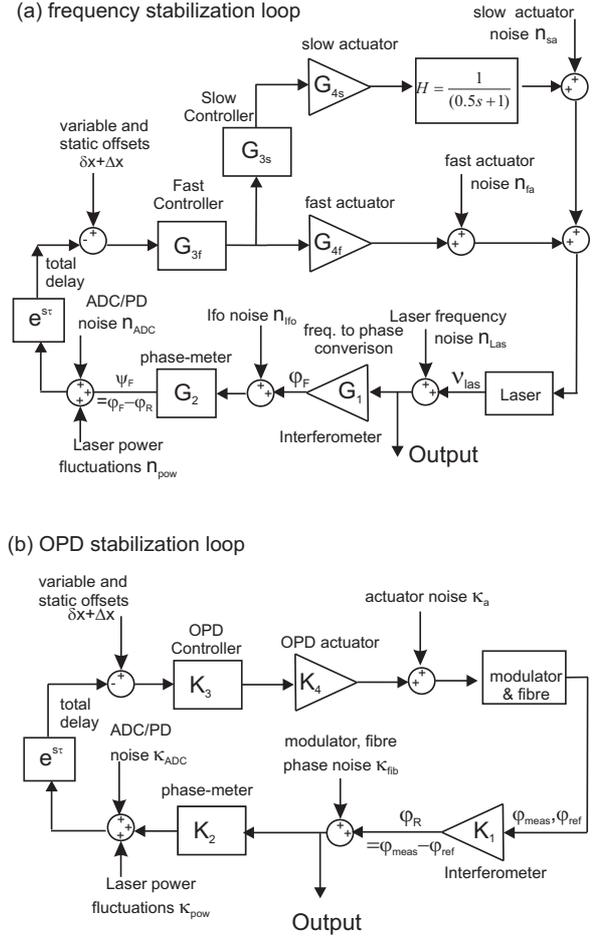

Figure 3: (a) A schematic of the laser frequency stabilization loops. The various gain stages and noise sources are denoted by $G_i$ and $n_i$ , respectively. The two loops are in cascaded configuration. Laser frequency fluctuations are converted into phase fluctuations through the frequency interferometer. (b) A schematic of the optical path-length difference stabilization loop. The various gain stages and noise sources are denoted by $K_i$ and $\kappa_i$ , respectively.

### D. The Optical path-length stabilization loop

A schematic of the optical path-length stabilization loop is given in Figure 3b. We start our discussion with the output of the laser modulator and fibre, at the top right corner. The phase of the measurement beam is denoted by $\varphi_{meas}$ and the phase of the reference beam by $\varphi_{ref}$ . Both phases fluctuate around their mean values due to phase drifts and fluctuations in fibre and modulator which induces noise $\kappa_{fib}$. The interferometer simply acts as a unity gain stage ($K_1 = 1$) for both phases and the phase difference $\varphi_R = \varphi_{meas} - \varphi_{ref}$ is detected in the interference pattern of the reference interferometer beating at the heterodyne frequency $f_{het}$.

The phase-meter converts the analogue phase into a digital and –as in the case of the frequency interferometer- adds ADC noise $\kappa_{ADC} = 7\,\mu rad/\sqrt{Hz}$ (smaller by a factor of $\sqrt{2}$ than $n_{ADC}$ because $\varphi_R$ is found from a single photo-diode unlike $\psi_F = \varphi_F - \varphi_R$, where two signals are subtracted) as well as noise induced by laser power fluctuations $\kappa_{pow}$.

Then the total loop delay $\tau_{OPD}$ is included in the schematic, followed by the input coupling of an optional dynamic operating setpoint $\Delta x + \delta x(t)$ right before the OPD loop controller with transfer function $K_3 = 4(s + 0.2)^2 / s^3$. The controller output goes to the actuator, a PZT controlled path-length actuator, with gain $K_4 \sim 6.6\,rad/V$. Here again some actuator noise is added which is estimated to be $\kappa_a < 350\,\mu rad/\sqrt{Hz}$ and, considering that it is suppressed by the total loop gain, can be neglected.

Taking into account all gain stages and transfer functions we obtain the following model for the open loop transfer function,

$$K_{tot} = K_4 \cdot \frac{4(s + 0.2)^2}{s^3} \cdot e^{s\tau}, \tag{5}$$

where we left the actuator gain $K_4$ and the total loop delay $\tau_{OPD}$ as free parameters.

## V. LOOP CHARACTERISTICS AND PERFORMANCE

### A. Measurement approach for the Open Loop Transfer Function

In order to determine the open loop transfer functions $G_{tot}$ and $K_{tot}$ of frequency and OPD loop, we apply a sinusoidal test signal $\delta x(t) = A \cdot \sin(2\pi\nu_{test}t)$ of frequency $\nu_{test}$ to the respective controller input. After waiting for some time for the transient response to die down we record a time series of at least 3 signal periods of the controlled quantity $x$ ($\psi_F$ or $\psi_R$) and of the control error $e$ (see Figure 2 for illustration). We then perform the discrete Fourier transform on these data and from the ratio of the complex amplitudes at the signal frequency we infer the value of the open loop transfer function at this particular frequency. Repeating the same measurement for a number of different signal frequencies, we obtain the characteristic function over the available control loop bandwidth. The measured and fitted open loop transfer functions are plotted in Figures 4 (a) and (b) below. The fits have been performed based on the model functions of Equations 4 and 5. We find excellent agreement between model and data which allows us to accurately determine the actuator gains from the fits.

### B. Frequency loop transfer function and noise model

For the frequency loop we find for the total loop delay $\tau_{OPD} = 41.1 \pm 1.0\,ms$ and for the slow and fast actuator gains $G_{4s} = 507 \pm 7\,MHz/V$ and $G_{4f} = 1.46 \pm 0.01\,MHz/V$, respectively. We additionally determine the fast frequency actuator gain in a separate open-loop measurement where we apply a ramp of given amplitude to the actuator and monitor the increase in phase $\psi_F$ in the frequency interferometer, from which we infer the gain to be $G_{4f} = 1.45 \pm 0.01\,MHz/V$. This is in excellent agreement with the gain extracted from the fit to the transfer function and demonstrates the accuracy of our loop modelling and characterization approach. As an additional bonus we also obtain the ratio of the path-length differences between frequency (path-length difference known to be $\sim 38\,cm$) and science interferometers, considering that the slope of the phase-change in both interferometers is proportional to the path-length difference, as suggested by Equation 2. The data imply that the ratio $r_{ad}$ of path-length differences between the frequency and the x1 interferometer is $r_{ad} = 39.0 \pm 0.7$ and between the frequency and the x12 interferometer $r_{ad} = -20.4 \pm 1.4$. Taking into account that the frequency interferometer has a path-length difference of 38 cm this result implies a path-length difference of approximately $1\,cm$ in the x1 and $-2\,cm$ in the x12 interferometer.

The unity gain frequency is found at $240\,mHz$ with a rather small phase margin of 27 degrees. From the transfer function we deduce that it would be easily possible to increase the overall loop gain by a factor of 10 and have enough phase-margin to stably operate the loop. The closed loop gain, given by the solid line in Figure 5, displays a pronounced "servo-bump" at unity gain due to the small phase-margin. At $30\,mHz$ the closed loop gain is $1.2 \times 10^{-2}$ and the free running laser noise is measured to be $\sim 400\,kHz/\sqrt{Hz}$, which results in an expected noise suppression to $\sim 5\,kHz/\sqrt{Hz}$, or equivalently $40\,\mu rad/\sqrt{Hz}$. As can be seen from the linear spectral density of the frequency fluctuations, plotted in the upper solid curve of Figure 6, this already corresponds to the asymptotic limit where the noise floor levels off. A further reduction in noise floor would be expected if the laser output noise was only given by the free running noise times the closed loop gain. This relation is given by the open white circles in Figure 6. Looking at the frequency loop schematic of Figure 3a we identify the likely origin of the limiting noise as the sensor noise which is not suppressed by the loop gain. However, the sensor noise is smaller than the asymptotic limit ($n_{ADC} \sim 10\,\mu rad/\sqrt{Hz}$).

The reason lies in the imperfect rejection of common mode noise which occurs in both, the phase of the frequency interferometer $\varphi_F$ and the phase of the reference interferometer $\varphi_R$. We found that the common mode noise rejection depends on the phase-relation between the two interferometers. In the subtraction of $\varphi_R$ from $\varphi_F$ the noise cancellation works optimally only if the phase difference is zero, i.e. $\psi_F = \varphi_F - \varphi_R = 0$. A simplified analytical model suggests that the noise rejection continuously deteriorates from $0 \deg$ to $90 \deg$ which is also indicated by the experimental data. For the data shown in Figure 6 we find that $\psi_F$ was locked to $\sim 100 \deg$ by the frequency servo which is close to worst case. However, even then the common mode noise rejection is good enough to easily fulfil the requirements. In another measurement where $\psi_F$ was locked to zero we find that the noise levels off at $10 \,\mu rad / \sqrt{Hz}$ at $1 \,mHz$ which corresponds to a frequency noise of $1.2 \,kHz / \sqrt{Hz}$. This is precisely the sensor noise predicted in [16] and gives further evidence of our assumptions on common mode noise rejection.

### C. OPD loop transfer function and noise model

For the OPD model we find a total loop delay of $\tau_{OPD} = 53.3 \pm 1.2 \,ms$ and an actuator gain of $K_4 = 4.88 \pm 0.06 \,rad / V$ by fitting the theoretical transfer function (Equation 5) to the measurement data. The loop delay is approximately $10 \,ms$ longer than for the frequency loop which fits with expectations that the higher OPD-loop filter order should increase the delay by one $100 \,Hz$ processing step. We also determine the actuator gain in a separate measurement where we apply a linear ramp over the whole actuation range and extract the gain as the ratio between measured phase-change to the applied ramp amplitude to find $K_4 \sim 8.40 \,rad / V$. The significant difference in value between this value and the one extracted from the transfer function fit can be attributed to the difference between small and large signal PZT gains, an effect which originates from PZT hysteresis and nonlinearities and has been observed in previous tests. The unity gain frequency is found at $3.1 \,Hz$ with a phase margin of $30 \deg$. Looking at the transfer function of Figure 4 we find that the total loop gain has already approached its limit and (provided we keep the chosen controller) cannot be increased any further without compromising loop stability. The closed loop gain, given by the dotted line in Figure 5, displays a pronounced "servo-bump" at unity gain due to the small phase-margin. In the performance measurement band it varies approximately $\propto f^3$ between $3.2 \times 10^{-7}$ (at $1 \,mHz$) and $4.6 \times 10^{-3}$ (at $30 \,mHz$).

From a measurement of the path-length noise in the laboratory under varying conditions we find that the open-loop OPD fluctuations are $\sim 900 \,rad / \sqrt{Hz}$ at $1 \,mHz$ and $\sim 20 \,rad / \sqrt{Hz}$ at $30 \,mHz$. The linear spectral density of the stabilized OPD fluctuations is plotted in the upper solid curve of Figure 6. We see that the measurement data agree well with a prediction based on the closed loop gain multiplied by the open loop fluctuations (given by the open triangles) between $1 \,mHz$ and $30 \,mHz$. Towards higher frequencies we measure a somewhat lower noise floor than predicted. This can be explained by our observation from a series of measurements where we found that there are rather large differences (up to $10 \,dB$) in the open loop OPD noise floor between $0.1 \,Hz$ and $1 \,Hz$.

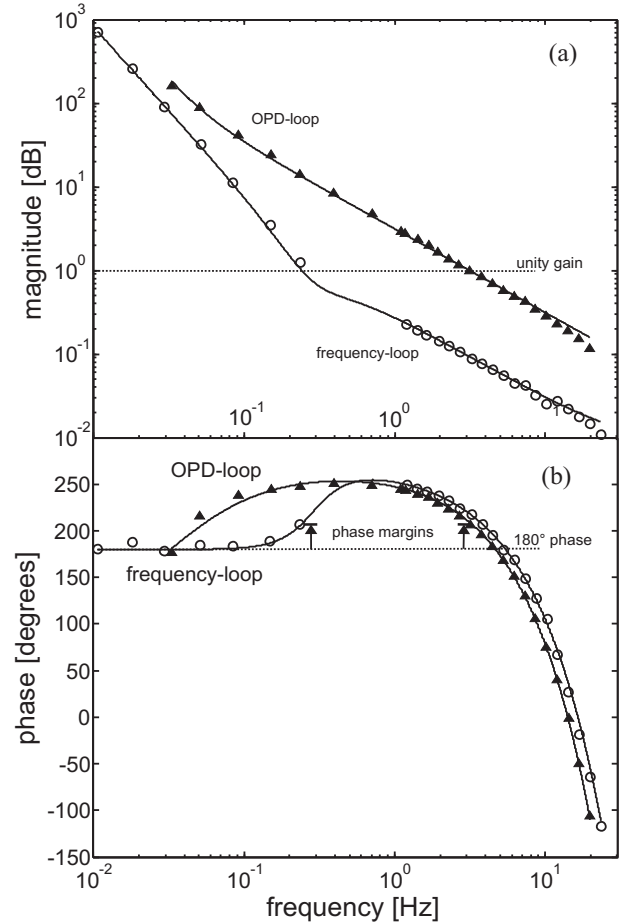

Figure 4: The open loop transfer function for the combination of fast and slow frequency loop (open circles) and the OPD loop (filled triangles) plotted in magnitude (a) and phase (b). A fit with a theoretical model function is given by the solid lines which have 3 (frequency loops) and 2 (OPD loop) free fitting parameters, respectively.

Towards lower frequency we find something rather surprising. Instead of being further suppressed as indicated by the prediction (asymptote indicated by dashed line), the measured closed loop OPD noise levels off at $700\,\mu rad/\sqrt{Hz}$ below $3\,mHz$. Referring to Figure 3b, we investigate the various noise contributions: The sensor noise $\kappa_{ADC} = 7\,\mu rad/\sqrt{Hz}$ is smaller by two orders of magnitude and therefore offers no explanation; neither does power noise $\kappa_{pow} \sim 1\,\mu rad/\sqrt{Hz}$ nor the actuator noise $\kappa_a < 350\,\mu rad/\sqrt{Hz}$ which is additionally suppressed by the closed loop gain of $\sim 10^{-5}$ at $3\,mHz$.

Recalling the basic working principle of the heterodyne interferometer we find that not only low frequency noise close to DC is measured by the interferometer but also high frequency noise around the heterodyne frequency $f_{het} = 1\,kHz$ couples into the measurement band.

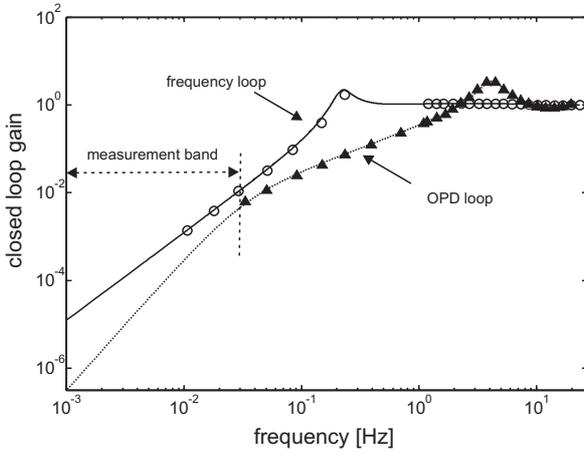

Figure 5: The magnitude of the closed-loop gain for the frequency loops (open circles) and the OPD loop (filled triangles) is plotted against frequency. The theoretical fits (derived from those of Figure 4) are given by the solid and dotted lines, respectively. The performance measurement band ($1\,mHz - 30\,mHz$) is indicated by the dashed arrows.

Considering that low frequency phase-noise should be strongly suppressed by the closed loop gain, we conclude that high frequency phase-noise is the origin of the phase-noise floor of $700\,\mu rad/\sqrt{Hz}$ below $3\,mHz$. This noise floor is not observed in the measurements of frequency (F) or science interferometers (x1,x12) from which we conclude that it must be common mode to all interferometers. We recall that the F,x1,x12 interferometers always subtract the reference phase $\varphi_R$ from their phase measurements and thereby cancel all common mode noise as well as possible. However, this noise is not common mode to the phases $\varphi_{meas}$ and $\varphi_{ref}$ of measurement and reference beam and therefore appears in the interference phase $\varphi_R$. This leads us to believe that the noise originates in the optical fibres or modulators of both beams. Note that by removing the MilBus converter and thereby reducing the loop delay would allow us to reduce the noise to approximately 1E-3 rad/Sqrt(Hz) in the measurement band between 3 and 30 mHz.

## D. Performance of the science interferometers

The science interferometers x1 and x12 measure the position and attitude of the two freely-floating test-masses which have been substituted for dummy mirrors in our ground-based test setup. The performance of the longitudinal position measurements for the x1 and x12 interferometer is given by the upper solid and dotted curve in Figure 7a, respectively. Both performance curves meet the requirement, indicated by the dashed line.

It is interesting to investigate the dependence of the overall performance on the contributions of the various noise sources. When we consider the impact of frequency noise we simply rescale the corresponding noise curve of Figure 6 by the ratio $r_{ad}$ of path-length differences between the science interferometer x1 and the frequency interferometer F.

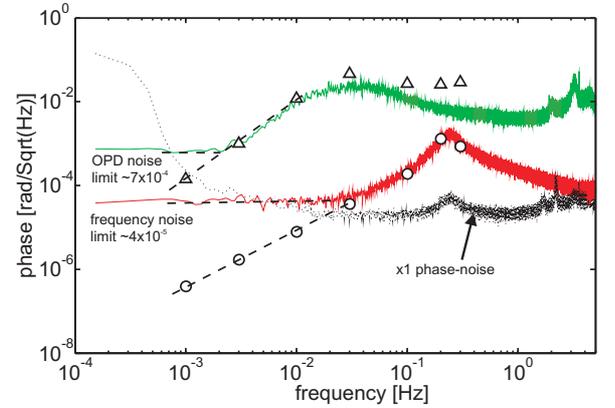

Figure 6: The phase-noise of the frequency loop (red solid line) and of the OPD loop (green solid line) are plotted against frequency. The expected phase noise for the two loops, defined as the open loop output noise times the closed loop gain, is given by the open circles and triangles, respectively. The asymptotic limits of the measured and expected noise curves are given by the dashed lines. For comparison the phase-noise of the science interferometer "x1" is given by the black dotted line.

In section 5B we found $r_{ad} = 39$ and therefore the noise contribution to the x1 performance is accordingly smaller by this factor. After converting the phase into position according to Equation 3 we obtain the frequency noise given by the lower dotted curve. It is apparent that frequency noise is completely negligible within the science measurement band and is only dominant very close to the frequency loop servo bump at $230\,mHz$ where we find perfect agreement between the measured x1 interferometer noise level and the predicted frequency noise level. We also notice a factor of 2 difference in the noise floor around the servo bump between the x1 and x12 interferometers which agrees perfectly with the fact that there is an path-length difference ratio of 2 between the science interferometers.

The OPD noise contribution to the science measurement is given by the product of the reference phase $\varphi_R$ times a factor related to the amplitude ratio between optical sideband and main carrier [17]. This factor was measured to be $\leq 60 \times 10^{-6}$ in preceding tests [18]. When we rescale the measured noise of the reference phase $\varphi_R$ by this factor and convert the resulting phase into position according to Equation 3, we obtain the lower solid line of Figure 7. Just like the frequency noise the OPD noise is much smaller than the measured science interferometer noise and can be completely neglected. The sensor noise limit corresponds to $\sim 0.88\, pm/\sqrt{Hz}$ [16] and is given by the dashed line in Figure 7, which is quite close to the measured noise levels of $\sim 2\, pm/\sqrt{Hz}$ for x1 and $\sim 3\, pm/\sqrt{Hz}$ for x12 at a frequency of $30\, mHz$.

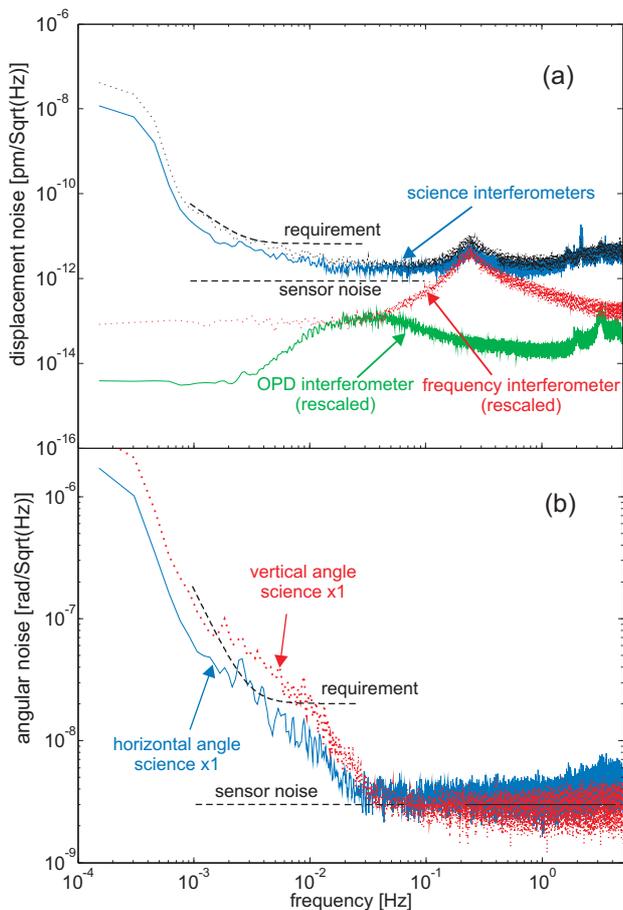

Figure 7: the performance of the science interferometers for longitudinal (a) and angular (b) measurements. Plot (a): the x1 longitudinal measurement performance is given by the blue solid line and the largely overlapping x12 performance by the black dotted line. The curved dashed line denotes the required system performance and the straight dashed line the sensor noise limit. The frequency noise contribution to the system performance is given by the red dotted line and the OPD noise contribution by the green solid line.

Plot (b): the horizontal angular measurement performance is given by the solid blue line, the vertical angular measurement performance by the dotted red line, and the required measurement performance by the curved dashed line. The lower dashed line represents the sensor limit.

We recall from section 5B our observation that common mode noise is not perfectly cancelled when the phase-difference is not equal to zero. As a consequence the frequency noise levelled off to $40\, \mu rad/\sqrt{Hz}$ instead of $10\, \mu rad/\sqrt{Hz}$ as expected from the sensor limit.

Similarly we find for the x1 and x12 interferometers that their mean differences to the reference phase $\varphi_R$ are $27\, \deg$ and $38\, \deg$, respectively, resulting in imperfect common mode noise rejection which is slightly worse for x1 compared with x12. The strong increase in noise after $\sim 3\, mHz$ might correspond to the temperature induced movements of the dummy-mirrors. A comparison between the noise curves of the science interferometers and the frequency interferometer (see Figure 6) suggests such a possibility. In the latter only the ultra-stable and bonded optical components of the interferometer are used and the noise level remains flat at low frequencies.

Figure 7b displays the angular performance of the horizontal angular measurement (solid line) and the vertical angular measurement (dotted line). Angular measurements do not determine the difference between the phases of two different interferometers, as is the case for the longitudinal phase measurements, but they are given by the phase difference between the diode quadrants of a single interferometer instead. The horizontal angle is obtained from the phase difference between the left and right half of the quadrant photo-diode and the vertical angle from the phase difference between upper and lower half of the quadrant diode [16,19].

The measured quadrant phase-differences $\theta$ are related to the test-mass tilt-angles $\vartheta$ through the coupling parameter $K : \theta = K \cdot \vartheta$. Although the latter depends on beam waist and wavefront radius of curvature as well as on the wavelength of light [19], the coupling parameters for both science interferometers are approximately given by $K \sim 5000$. Dividing through that factor we convert the measured phase-noise into angular phase-noise of the test-mass. The required accuracy for angular measurements is given by the dashed line and we find that it is violated for the vertical angular measurements below $10\, mHz$. There is a clear difference in noise level between the vertical and horizontal angular measurements for frequencies below $30\, mHz$. While we are not certain about its origin, a possible explanation could be thermal fluctuations which -through the way the fibre injectors are mounted- couples preferentially in the vertical direction.

For comparison, the predicted sensor noise limit of $3 \times 10^{-9}\, rad/\sqrt{Hz}$ [16], is plotted as the dashed line and we find perfect agreement with the measurement data for frequencies above $30\, mHz$. This is clear evidence that the system is only sensor-noise limited for higher frequencies.

## VI. CONCLUSIONS

We report on the first investigations of control loops and performance measurements with the complete engineering model of the Optical Metrology System (OMS) to fly on-board the LISA Pathfinder mission to space.

The loop architecture is described and analyzed, and detailed noise models of the loops are introduced and compared with the results of performance measurements. The theoretically derived open loop transfer functions accurately fit the measurement data and allow us to determine actuator gains and loop delays. The accuracy of the "science measurement" of longitudinal test-mass position was shown to be close to the sensor-noise limit and well below the requirements. We also showed that the cancellation of common mode noise, which is shared by all four interferometers, works best if the relative phases are zero, but the required measurement performance is met even if this condition is not fulfilled.

Both, frequency and OPD control loop were shown to experience large loop delays. These are dominated by the contribution of an interface converter (serial to MIL-STD-1553) that is not present in the actual flight model of the system, which should therefore significantly improve the achievable loop gains.

### Acknowledgment

We gratefully acknowledge the financial support by Deutsches Zentrum für Luft- und Raumfahrt (DLR) to perform the experiments and measurements with the optical metrology system described herein on the premises of the Albert Einstein Institute. We would like to thank all personnel of the institute and of Astrium who contributed. We would also like to thank D. Fertin and M. Cesa of the European Space Agency (ESA), J. Reiche from Albert Einstein Institut, and P. Bergner, R. Gerndt and U. Johann from Astrium, for their help and many useful discussions.